\documentclass[11pt]{revtex4}
\usepackage{amsmath}
\usepackage{amssymb,epsf}
\usepackage{latexsym}
\usepackage{amssymb}
\usepackage{amsmath}
\usepackage{graphicx}
\usepackage{subfigure}
\usepackage{epstopdf}

\begin{document}

\title{Counterterms for Static Lovelock Solutions}
\author{M. R. Mehdizadeh$^{1}$, M. H. Dehghani$^{2,3}$, and M. Kord Zangeneh$%
^{3}$}
\affiliation{$^1$ Department of Physics, Shahid Bahonar University, PO Box 76175, Kerman
76175, Iran\\
$^2$Research Institute for Astrophysics and Astronomy of Maragha (RIAAM),
Maragha, Iran\\
$^3$Physics Department and Biruni Observatory, College of Sciences, Shiraz
University, Shiraz 71454, Iran}

\begin{abstract}
In this paper, we introduce the counterterms that remove the non-logarithmic
divergences of the action in third order Lovelock gravity for static
spacetimes. We do this by defining the cosmological constant in such a way
that the asymptotic form of the metric have the same form in Lovelock and
Einstein gravities. Thus, we employ the counterterms of Einstein gravity and
show that the power law divergences of the action of Lovelock gravity
for static spacetimes can be removed by suitable choice of
coefficients. We find that the dependence of these coefficients on the
dimension in Lovelock gravity is the same as in Einstein gravity. We also
introduce the finite energy-momentum tensor and employ these counterterms to
calculate the finite action and mass of static black hole solutions
of third order Lovelock gravity. Next, we calculate the thermodynamic
quantities and show that the entropy calculated through the use of
Gibbs-Duhem relation is consistent with the obtained entropy by Wald's
formula. Furthermore, we find that in contrast to Einstein gravity in which
there exists no uncharged extreme black hole, third order Lovelock gravity
can have these kind of black holes. Finally, we investigate the stability of
static charged black holes of Lovelock gravity in canonical
ensemble and find that small black holes show a phase transition between
very small and small black holes, while the large ones are stable.
\end{abstract}

\maketitle

\section{Introduction}

An interesting framework for studying the non-perturbative quantum field
theories is through the use of anti-de Sitter/conformal field theory
(AdS/CFT) correspondence \cite{Mal}. According to this duality, in
principle, one can perform gravity calculations to find information about
the field theory side or vice versa. In this context, the central charges of
the dual theory (CFT) relate to coupling constants of its dual gravity.
Therefore, Einstein gravity with one coupling constant restricts the dual
theory to a limited class of CFT with equal central charges \cite{Ein}. For
extension of the duality beyond this limit, one needs to involve higher
curvature terms in the gravity action. It is clear that each correction term
introduces a new coupling constant and therefore one may have CFT theory
with different central charges. Indeed, this procedure leads to the richness
of the CFT theory \cite{HolLov}. The most natural extension of general
relativity with higher curvature terms and with the assumption of Einstein
-- that the left hand side of the field equations is the most general
symmetric conserved tensor containing no more than two-derivatives of the
metric -- is Lovelock theory. Lovelock \cite{Love} found the most general
symmetric conserved tensor satisfying this property. The resultant tensor is
nonlinear in the Riemann tensor and differs from the Einstein tensor only if
the spacetime has more than 4 dimensions. Although Lovelock gravity leads to
second-order field equations and it has ghost free AdS solution \cite{ghost}%
, it has been recently shown that quadratic and cubic gravities entail
causality violation and there are stringent conditions on the coupling
constants \cite{Edel}.

The problem with the total action of Einstein gravity is that it is
divergent when evaluated on the solutions \cite{BY,BCM,BoothMann}. Due to
this fact, all the other conserved quantities which is calculated through
the use of this action is also divergent. One way of eliminating these
divergences is through the use of background subtraction method of Brown and
York \cite{BY}. In this method, the boundary surface is embedded in another
(background) spacetime, and one subtracts the action evaluated on the
embedded surface of the background spacetime from the total action. Such a
procedure causes the resulting physical quantities to depend on the choice
of reference background. Furthermore, it is not possible in general to embed
the boundary surface into a background spacetime. For asymptotically AdS
solutions of Einstein gravity, one may remove the non-logarithmic
divergences in the action by adding a counterterm action which is a
functional of the boundary curvature invariants \cite{Kraus,Emp}. Indeed,
this counterterm method furnishes a means for calculating the action and
conserved quantities intrinsically without reliance on any reference
spacetime \cite{Sken,BK,Od2}. Although there may exist a very large number
of possible invariants, only a finite number of them are non-vanishing in a
given dimension on a boundary at infinity. This method has been applied to
many cases such as black holes with rotation, NUT charge, various
topologies, rotating black strings with zero curvature horizons and rotating
higher genus black branes \cite{Deh3}. Although the counterterm method
applies for the case of a specially infinite boundary, it was also employed
for the computation of the conserved and thermodynamic quantities in the
case of a finite boundary \cite{DM2}.

All of the works mentioned in the previous paragraph were limited to
Einstein gravity. Although the counterterm of Lovelock gravity with flat
horizon has been introduced \cite{Deh1, ShD}, only a few works related to
the counterterm method have been done for Lovelock gravity with curved
horizon. This is due to the fact that even for Einstein gravity, the
systematic construction that provides the form of the counterterms becomes
cumbersome for high enough dimensions \cite{Kraus,Emp}. Indeed in this
method, one should reconstruct the spacetime metric by solving iteratively
the field equations in the Fefferman-Graham frame \cite{Fef}:
\begin{equation}
ds^{2}=\frac{l^{2}}{4\rho ^{2}}d\rho ^{2}+\frac{1}{\rho }\left[
g_{(0)ij}(x)+\rho g_{(1)ij}(x)+\rho ^{2}g_{(2)ij}(x)+...\right] dx^{i}dx^{j},
\label{FefMet}
\end{equation}%
where $\mathbf{g}_{\mathbf{(0)ij}}(x)$ is the boundary data of an
initial-value problem governed by the equations of motion. However, even for
Einstein-Hilbert theory, solving the coefficients $\mathbf{g}_{(p)}(x)$ in
Eq. (\ref{FefMet}) as covariant functionals of $\mathbf{g}_{(0)}(x)$ is only
possible for low enough dimensions. Thus, it is expected that the
holographic renormalization procedure would be even more complicated in
Lovelock gravity because of the nonlinearity of field equations. Indeed,
because of the nonlinearity of the field equations solving $\mathbf{g}%
_{(p)}(x)$ in Eq. (\ref{FefMet}) as covariant functionals of $\mathbf{g}%
_{(0)}(x)$ would be even more cumbersome. So, the authors in Ref. \cite{Olea}
presented an alternative construction of Kounterterms. Instead of adding
counterterms to cancel the divergence at the boundary explained above, they
circumvented the difficulties of the standard method by using Kounterterms
which depend on the intrinsic and the extrinsic curvatures of the boundary.
They selected the Kounterterms as the boundary terms which are regular on
the asymptotic region. Indeed, the regularization process is encoded in the
boundary terms already presented and there is no need to add further
counterterms.

The exact rotating solutions of Lovelock gravity with curved horizon
are not introduced till now. Indeed, only static solutions of Lovelock
gravity with different matter fields are known \cite{Stat}. So, because of
the difficulties of the holographic renormalization procedure in Lovelock
gravity and the nonexistance of an exact rotating solution of this theory,
we limit ourselves to the case of counterterms of Lovelock gravity for
static solutions. The counterterms of asymptotically AdS static solutions of
Gauss-Bonnet gravity have been introduced in Ref. \cite{Radu, Yale}. Also,
the finite action and global charges of asymptotically de Sitter static
solutions has been obtained in Ref. \cite{Radu2}.

Here we like to apply the counterterm method to the case of the 
static solutions of the field equations of third order Lovelock gravity
with curved horizon. We define the cosmological constant in such a way that
the maximally symmetric AdS spacetime
\begin{equation}
ds^{2}=-\left( k+\frac{r^{2}}{L^{2}}\right) dt^{2}+\left( k+\frac{r^{2}}{%
L^{2}}\right) ^{-1}dr^{2}+r^{2}d\Sigma _{k,n-1}^{2}  \label{AdSmet}
\end{equation}%
be the vacuum solution of Lovelock gravity. In Eq. (\ref{AdSmet}) $d\Sigma
_{k,n-1}^{2}$ is the metric of an ($n-1$)-dimensional maximally symmetric
space with curvature constant $(n-1)(n-2)k$ and volume $V_{k,n-1}$. Indeed,
this choice of cosmological constant makes the asymptotic form of the
solutions of Lovelock gravity to be exactly the same as that of Einstein
gravity. Thus, we expect that the counterterm introduced for Einstein
gravity in \cite{Kraus} may remove the power law divergences in the action
of Lovelock gravity. Although the counterterms which should be added to
Gauss-Bonnet gravity in order to remove the power law divergences of the
action for static solutions are introduced in Ref. \cite{Radu},
they depend on the Gauss-Bonnet coefficient. However, because of our choice
of the cosmological constant, our counterterms are the same as those of
Einstein gravity and are independent of Lovelock coefficients. In order to
check our counterterms, we calculate the finite action and the mass of the
black hole through the use of counterterm method. Then, we use these finite
quantities and the Gibbs-Duhem relation to obtain the entropy. We find that
the calculated entropy of the black holes is consistent with the Wald's
formula \cite{Wald}. As another test of our counterterm method, we show that
the mass obtained through the use of counterterm method satisfies the first
law of thermodynamics. We, also, perform a stability analysis of the black
hole solutions in canonical ensemble and investigate the effects of third
order Lovelock term on the stability.

This paper is organized as follows. In section \ref{Sol}, we review the
well-defined action of Lovelock gravity. In section \ref{Count}, we
introduce the counterterms for third order Lovelock gravity for
static spacetimes. We also, introduce the finite stress energy tensor of
this theory. Section \ref{Therm} is devoted to the thermodynamics of the
black hole solutions of the theory. We calculate the finite action, the
total mass, the temperature, the charge and the electric potential. We
calculate the entropy through the use of Gibbs-Duhem relation and Wald
formula and find that they are consistent. We, also, investigate the first
law of thermodynamics. In Sec. \ref{Stab}, we investigate the thermal
stability of the solutions in canonical ensemble. We finish our paper with
some concluding remarks.

\section{Action and Field equations\label{Sol}}

The bulk action of Lovelock gravity in $n+1$ dimensions may be written as
\cite{Love}
\begin{equation}
I_{bulk}=\frac{1}{16\pi }\int_{\mathcal{M}}dx^{n+1}\sqrt{-g}\sum_{p=1}^{%
\left[ n/2\right] }\alpha _{p}\left( \mathcal{L}_{p}-2\Lambda _{p}\right)
+I_{mat}  \label{action1}
\end{equation}%
with $[x]$ denoting the integer part of $x$, $\alpha _{p}$'s ($p\geq 2$) are
Lovelock coefficients,
\begin{equation*}
\mathcal{L}_{p}=\frac{1}{2^{p}}\delta
_{b_{1}b_{2}...b_{2p-1}b_{2p}}^{a_{1}a_{2}...a_{2p-1}a_{2p}}R_{a_{1}a_{2}}^{b_{1}b_{2}}...R_{a_{2p-1}a_{2p}}^{b_{2p-1b_{2p}}}
\end{equation*}%
is the Euler density of a $2p$-dimensional manifold, $\delta
_{b_{1}...b_{2p}}^{a_{1}...a_{2p}}$ is the general asymmetric kronecker
delta and%
\begin{equation}
\Lambda _{p}=\frac{(-1)^{p}n(n-1)(n-2)\cdots (n-2p+1)}{2L^{2p}}.  \label{Cos}
\end{equation}%
The action (\ref{action1}) is written in such a way that the maximally
symmetric AdS spacetime (\ref{AdSmet}) is the vacuum solution of action (\ref%
{action1}). In this notation, the independent coupling constants are $L$ and
all the Lovelock coefficients.

From a geometric point of view the Lagrangian of the action (\ref{action1})
in $2[n/2]+1$ and $2[n/2]+2$ dimensions is the most general Lagrangian that
yields second order field equations, as in the case of Einstein-Hilbert
action which is the most general Lagrangian producing second order field
equations in three and four dimensions. In the rest of the paper, we work in
a unit system with $\alpha _{1}=\tilde{\alpha}_{1}=1$ and the dimensionless
Lovelock coefficients $\tilde{\alpha}_{p}$ defined as%
\begin{equation}
\tilde{\alpha}_{p}\equiv \frac{(n-2)...(n-2p+1)}{L^{2(p-1)}}{\alpha _{p},}%
\text{ \ }p\geq 2.  \label{alphahat}
\end{equation}%
With the definition (\ref{alphahat}), the cosmological constant for AdS
spacetime is%
\begin{equation*}
\Lambda =\frac{n(n-1)}{2L^{2}}\sum_{p=1}^{\left[ n/2\right] }(-1)^{p}\tilde{%
\alpha}_{p}.
\end{equation*}

In this paper, we consider the third order Lovelock gravity in the presence
of electromagnetic field. Thus, the action of matter field is
\begin{equation*}
I_{mat}=-\frac{1}{64\pi }\int_{\mathcal{M}}dx^{n+1}\sqrt{-g}\partial
_{\lbrack \mu }A_{\nu ]}\partial ^{\lbrack \mu }A^{\nu ]},
\end{equation*}%
where $A_{\mu }$ is the electromagnetic potential. The first term in
Lovelock Lagrangian is the Einstein-Hilbert term $R$, the second term is the
Gauss-Bonnet Lagrangian $\mathcal{L}_{2}=R_{\mu \nu \gamma \delta }R^{\mu
\nu \gamma \delta }-4R_{\mu \nu }R^{\mu \nu }+R^{2}$, the third term is
\begin{eqnarray}
\mathcal{L}_{3} &=&2R^{\mu \nu \sigma \kappa }R_{\sigma \kappa \rho \tau }R_{%
\phantom{\rho \tau }{\mu \nu }}^{\rho \tau }+8R_{\phantom{\mu \nu}{\sigma
\rho}}^{\mu \nu }R_{\phantom {\sigma \kappa} {\nu \tau}}^{\sigma \kappa }R_{%
\phantom{\rho \tau}{ \mu \kappa}}^{\rho \tau }+24R^{\mu \nu \sigma \kappa
}R_{\sigma \kappa \nu \rho }R_{\phantom{\rho}{\mu}}^{\rho }  \notag \\
&&+3RR^{\mu \nu \sigma \kappa }R_{\sigma \kappa \mu \nu }+24R^{\mu \nu
\sigma \kappa }R_{\sigma \mu }R_{\kappa \nu }+16R^{\mu \nu }R_{\nu \sigma
}R_{\phantom{\sigma}{\mu}}^{\sigma }-12RR^{\mu \nu }R_{\mu \nu }+R^{3},
\label{L3}
\end{eqnarray}%
and the cosmological constant is%
\begin{equation}
\Lambda =-\frac{n(n-1)}{2L^{2}}(1-\tilde{\alpha}_{2}+\tilde{\alpha}_{3}).
\label{Lam3}
\end{equation}

As in the case of Einstein-Hilbert action, the action (\ref{action1}) does
not have a well-defined variational principle, since one encounters a total
derivative that produces a surface integral involving the derivatives of $%
\delta g_{\mu \nu }$ normal to the boundary $\partial \mathcal{M}$. These
normal derivatives of $\delta g_{\mu \nu }$ can be canceled by the variation
of the surface action \cite{Myers1,Deh1}
\begin{equation*}
I_{sur}=\frac{1}{8\pi }\int_{\partial \mathcal{M}}d^{n+1}x\sqrt{-\gamma }%
\sum_{p=1}^{\left[ n/2\right] }\alpha _{p}Q_{p},
\end{equation*}%
where
\begin{eqnarray}
Q_{p} &=&p\int\limits_{0}^{1}dt\,\delta _{\lbrack ii_{1}\cdots
i_{2p-1}]}^{[jj_{1}\cdots j_{2p-1}]}\,K_{j_{1}}^{i_{1}}\,\times  \notag \\
&&\times \left( \frac{1}{2}\,\hat{R}_{j_{2}j_{3}}^{i_{2}i_{3}}(\gamma
)-t^{2}K_{j_{2}}^{i_{2}}K_{j_{3}}^{i_{3}}\right) \cdots \left( \frac{1}{2}\,%
\hat{R}_{j_{2p-2}j_{2p-1}}^{i_{2p-2}i_{2p-1}}(\gamma
)-t^{2}\,K_{j_{2p-2}}^{i_{2p-2}}K_{j_{2p-1}}^{i_{2p-1}}\right) .
\label{SurLag}
\end{eqnarray}%
In Eq. (\ref{SurLag}) $\gamma _{ab}$ and $K_{ab}=-\gamma _{a}^{\mu }\nabla
_{\mu }n_{b}$ are the induced metric and extrinsic curvature of the boundary
$\partial \mathcal{M}$, respectively. The explicit form of the first three
terms of Eq. (\ref{SurLag}) are \cite{Deh1}:
\begin{eqnarray}
I_{sur}^{(1)} &=&\frac{1}{8\pi }\int_{\partial \mathcal{M}}d^{n}x\sqrt{%
-\gamma }K~,  \notag \\
I_{sur}^{(2)} &=&\frac{\alpha _{2}}{4\pi }\int_{\partial \mathcal{M}}d^{n}x%
\sqrt{-\gamma }\left( J-2\hat{G}_{ab}^{(1)}K^{ab}\right) ~,  \notag \\
I_{sur}^{(3)} &=&\frac{3\alpha _{3}}{8\pi }\int_{\delta \mathcal{M}}d^{n}x%
\sqrt{-\gamma }\left\{ P-2\hat{G}_{ab}^{(2)}K^{ab}+2\hat{R}J\right.  \notag
\\
&&\left. -12\hat{R}_{ab}J^{ab}-4\hat{R}_{abcd}\left(
2K^{ac}K_{e}^{b}K^{ed}-KK^{ac}K^{bd}\right) \right\} ,  \label{Isurf}
\end{eqnarray}%
where $J$ and $P$ are the traces of
\begin{equation*}
J_{ab}=\frac{1}{3}%
(2KK_{ac}K_{b}^{c}+K_{cd}K^{cd}K_{ab}-2K_{ac}K^{cd}K_{db}-K^{2}K_{ab})
\end{equation*}%
and
\begin{eqnarray}
P_{ab} &=&\frac{1}{5}%
\{[K^{4}-6K^{2}K^{cd}K_{cd}+8KK_{cd}K_{e}^{d}K^{ec}-6K_{cd}K^{de}K_{ef}K^{fc}+3(K_{cd}K^{cd})^{2}]K_{ab}
\notag \\
&&-(4K^{3}-12KK_{ed}K^{ed}+8K_{de}K_{f}^{e}K^{fd})K_{ac}K_{b}^{c}-24KK_{ac}K^{cd}K_{de}K_{b}^{e}
\notag \\
&&+(12K^{2}-12K_{ef}K^{ef})K_{ac}K^{cd}K_{db}+24K_{ac}K^{cd}K_{de}K^{ef}K_{bf}\},
\label{Pab}
\end{eqnarray}%
respectively. In Eq. (\ref{Isurf}) $\hat{G}_{ab}^{(1)}$ is the Einstein
tensor, $\hat{R}_{abcd}(\gamma )$ is the intrinsic curvature and $\hat{G}%
_{ab}^{(2)}$ is the Gauss-Bonnet tensor of the metric $\gamma _{ab}$ given
as
\begin{equation*}
\hat{G}_{ab}^{(2)}=2(\hat{R}_{acde}\hat{R}_{b}^{\phantom{b}{c d e}}-2\hat{R}%
_{acbd}\hat{R}^{cd}-2\hat{R}_{ac}\hat{R}_{\phantom{c}b}^{c}+\hat{R}\hat{R}%
_{ab})-\frac{1}{2}\mathcal{\hat{L}}_{2}\gamma _{ab}.
\end{equation*}

\section{Counterterm method for static solutions of third order Lovelock
gravity \label{Count}}

It is well known that the action $I_{bulk}+I_{sur}$ is not finite for
asymptotically AdS solutions. Inspired by AdS/CFT correspondence, one needs
to add counterterms to the gravity action in order to get a finite action.
These counterterms are made from the curvature invariants of the boundary
metric with the coefficients of the higher curvature terms chosen so that
power law divergences in the bulk are canceled for all possible boundary
topologies permitted by the equations of motion. At any given dimension
there are only a finite number of counterterms that do not vanish at
infinity. This does not depend upon what the gravity theory is -- i.e.
whether or not it is Einstein, Gauss-Bonnet, 3rd order Lovelock, etc.
Indeed, for asymptotically AdS solutions, the boundary counterterms that
cancel the divergences in Einstein Gravity may also cancel the divergences
in Lovelock gravity if one chooses the cosmological constant as in Eq. (\ref%
{Cos}). This is due to the fact that the $p$th order Lovelock Lagrangian $%
\sqrt{\gamma }\mathcal{L}_{p}$ calculated for the metric (\ref{AdSmet}) is
independent of Lovelock coefficients. That is, the different orders of
Lovelock action do not mix with each other and one may find the counterterms
for different orders of Lovelock terms separately. This point makes the
calculation easier. Of course, the coefficients of the various counterterms
for different Lovelock terms will be different, depend only on $L$ and will
be independent of Lovelock coefficients. Thus, using the counterterms of
Einstein gravity \cite{Kraus}, we may write the counterterms of third order
Lovelock gravity as
\begin{equation}
I_{ct}=\sum_{p=1}^{3}I_{ct}^{(p)},  \label{Counter}
\end{equation}%
where
\begin{eqnarray}
I_{ct}^{(p)} &=&\frac{1}{8\pi }\int_{\partial M}dx^{n}\sqrt{-\gamma }\tilde{%
\alpha}_{p}\left\{ A_{p}+B_{p}\hat{R}+C_{p}\left( \hat{R}^{ab}\hat{R}_{ab}-%
\frac{n}{4(n-1)}\hat{R}^{2}\right) \right.  \notag \\
&&\left. +D_{p}\frac{3n+2}{4(n-1)}\hat{R}\hat{R}^{ab}\hat{R}_{ab}-\frac{%
n\left( n+2\right) }{16(n-1)^{2}}\hat{R}^{3}-2\hat{R}^{ab}\hat{R}^{cd}\hat{R}%
_{acbd}\right.  \notag \\
&&\left. +\frac{n-2}{2(n-1)}\hat{R}^{ab}D_{a}D_{b}\hat{R}-\hat{R}^{ab}D^{2}%
\hat{R}_{ab}+\frac{1}{2(n-1)}\hat{R}D^{2}\hat{R}+\cdots \right\} .
\label{Ict-p}
\end{eqnarray}%
The coefficients $A_{p}$, $B_{p}$, $C_{p}$ and $D_{p}$ in Eq. (\ref{Ict-p})
should depend on $L$. These coefficients for Einstein gravity ($\tilde{\alpha%
}_{p}=1$, $\tilde{\alpha}_{2}=\tilde{\alpha}_{3}=0$) are \cite{Kraus}
\begin{equation}
A_{1}=-\frac{(n-1)}{L},\quad B_{1}=-\frac{L}{2(n-2)},\quad C_{1}=-\frac{L^{3}%
}{2(n-2)^{2}(n-4)},\quad D_{1}=\frac{L^{5}}{(n-2)^{3}(n-4)(n-6)}  \label{A1}
\end{equation}%
One may note that the coefficients $A_{p}$, $B_{p}$, $C_{p}$ and $D_{p}$ are
independent of Lovelock coefficients. We apply the counterterms (\ref{Ict-p}%
) to various static solutions of Gauss-Bonnet and third order
Lovelock gravity with different topolgy and find that these coefficients are
\begin{equation}
A_{2}=-\frac{2}{3}A_{1},\quad B_{2}=B_{1},\quad C_{2}=-6C_{1},\text{ \ \ \ \
\ \ }D_{2}=-10D_{1},  \label{A2}
\end{equation}%
and
\begin{equation}
A_{3}=\frac{6}{5}A_{1},\text{ \ \ \ }B_{3}=-B_{1},\text{ \ \ }%
C_{3}=-18C_{1},\qquad D_{3}=15D_{1},  \label{A3}
\end{equation}%
for Gauss-Bonnet and third order Lovelock gravity, respectively. The
reason that we use exactly the counterterms of Einstein gravity is as
follows. First, the boundary at $\mathbf{r=const.}$ for static
solutions is a constant curvature hypersurface and therefore no
six-derivative term will be appeared in the counterterms. In other words,
all the terms of a specific order of counterterms [for example $R^{2}$
and $R_{ab}R^{ab}$ in $C_{2}( \hat{R}^{ab}\hat{R}_{ab}-n%
\hat{R}^{2}/[4(n-1)]) $] for static solutions are proportional to $%
r^{-4}$. Also, $R_{abcd}R^{abcd}$ is proportional to $%
r^{-4}$. Therefore, in order to remove the divergences of the action
which are proportional to $r^{-4}$, any combination of $R^{2}$, 
$R_{ab}R^{ab}$ and $R_{abcd}R^{abcd}$ can be
used. So, we just use exactly the counterterms of Einstein gravity. Second, 
$A_{p}$\textbf{, }$B_{p}$\textbf{, }$C_{p}$\textbf{, }$D_{p}$, for $%
p=2$ and $3$ are proportional to those of Einstein
counterterm independent of the dimensions. That is, the
dimensional-dependence of these coefficients are the same as those in
Einstein gravity. Third, as we will see in the next section, the entropy
calculated through the use of Gibbs-Duhem relation and the mass and action
calculated by our counterterms is consistent with the entropy obtained by
use of Wald's formula. Fourth, the mass calculated by our counterterms
satisfies the first law of thermodynamics.

While the total action $I_{bulk}+I_{sur}+I_{ct}$ is appropriate in
grand-canonical ensemble where $\delta A_{\mu }$ is zero at the boundary,
the appropriate action in the canonical ensemble where the electric charge
is fixed is \cite{Haw}
\begin{equation}
\tilde{I}=I_{bulk}+I_{sur}+I_{ct}-\frac{1}{16\pi }\int_{\partial \mathcal{M}%
}d^{n}x\sqrt{-\gamma }n^{\mu }F_{\mu \nu }A^{\nu }.  \label{ActCan}
\end{equation}%
Thus both in canonical and grand-canonical ensemble, the variation of total
action about the solutions of the field equations is
\begin{equation}
\delta I=\delta \tilde{I}=\sum_{p=0}^{3}\left( \frac{\delta I_{sur}^{(p)}}{%
\delta \gamma ^{ab}}+\frac{\delta I_{ct}^{(p)}}{\delta \gamma ^{ab}}\right)
\delta \gamma ^{ab}.
\end{equation}%
So, the energy-momentum tensor can be written as:
\begin{equation}
T_{ab}=T_{ab}^{(sur)}+T_{ab}^{(ct)}=\frac{2}{\sqrt{-\gamma }}%
\sum_{p=0}^{3}\left( \frac{\delta I_{sur}^{(p)}}{\delta \gamma ^{ab}}+\frac{%
\delta I_{ct}^{(p)}}{\delta \gamma ^{ab}}\right) .  \label{Stres}
\end{equation}%
The explicit expressions of $T_{ab}^{(sur)}$ and $T_{ab}^{(ct)}$ are
somewhat cumbersome so we give them in the Appendix.

To compute the conserved charges of the spacetime, we choose a spacelike
hypersurface $\mathcal{B}$ in $\partial \mathcal{M}$ with metric $\sigma
_{ij}$, and write the boundary metric in ADM form:
\begin{equation}
\gamma _{ab}dx^{a}dx^{a}=-N^{2}dt^{2}+\sigma _{ij}\left( d\varphi
^{i}+V^{i}dt\right) \left( d\varphi ^{j}+V^{j}dt\right) ,
\end{equation}%
where the coordinates $\varphi ^{i}$ are the angular variables
parameterizing the hypersurface of constant $r$ around the origin, and $N$
and $V^{i}$ are the lapse and shift functions, respectively. When there is a
Killing vector field $\varsigma $ on the boundary, then the quasilocal
conserved quantities associated with the stress tensors of Eq. (\ref{Stres})
can be written as
\begin{equation}
\mathcal{Q}(\varsigma)=\int_{\mathcal{B}}d^{n-1}\varphi \sqrt{\sigma }%
T_{ab}n^{a} \varsigma^{b},  \label{charge}
\end{equation}%
where $\sigma $ is the determinant of the metric $\sigma _{ij}$, and $n^{a}$
is the timelike unit normal vector to the boundary $\mathcal{B}$. In the
context of counterterm method, the limit in which the boundary $\mathcal{B}$
becomes infinite ($\mathcal{B}_{\infty }$) is taken, and the counterterm
prescription ensures that the action and conserved charges are finite. No
embedding of the surface $\mathcal{B}$ into a reference of spacetime is
required and the quantities which are computed are intrinsic to the
spacetimes.

\section{Thermodynamics of AdS Charged Black Holes\label{Therm}}

The field equation of third order Lovelock gravity for the static metric
\begin{equation*}
ds^{2}=-f(r)dt^{2}+\frac{dr^{2}}{f(r)}+r^{2}d\Sigma _{k,n-1}^{2}
\end{equation*}%
in the presence of electromagnetic field may be written as
\begin{equation}
\tilde{\alpha}_{3}([\psi ^{3}(r)-1]-\tilde{\alpha}_{2}[\psi ^{2}(r)-1]+[\psi
(r)-1]+\frac{mL^{2}}{r^{n}}-\frac{2}{(n-1)(n-2)}\frac{q^{2}L^{2}}{r^{2(n-1)}}%
=0,  \label{Cubic}
\end{equation}%
where $\psi (r)\equiv \lbrack f(r)-k]L^{2}/r^{2}$, $q$ is the charge of the
black hole and $m$ is the integration constant which is related to the mass
of the solution. The electromagnetic potential for the above metric is%
\begin{equation*}
A_{\mu }=-\frac{q}{(n-2)r^{n-2}}\delta _{\mu }^{t}.
\end{equation*}%
The mass parameter $m$ may be written in terms of horizon radius and $q$ by
using the fact that $\psi (r_{+})=-kL^{2}/r_{+}^{2}$, and therefore
\begin{equation*}
m=\left( 1-\tilde{\alpha}_{2}+\tilde{\alpha}_{3}+k\frac{L^{2}}{r_{+}^{2}}\mu
\right) \frac{r_{+}^{n}}{L^{2}}+\frac{2q^{2}}{(n-1)(n-2)r_{+}^{n-2}},
\end{equation*}%
where%
\begin{equation}
\mu =1+\tilde{\alpha}_{2}k\frac{L^{2}}{r_{+}^{2}}+\tilde{\alpha}_{3}k^{2}%
\frac{L^{4}}{r_{+}^{4}}.  \label{mu}
\end{equation}%
As one expects, $\psi (r)=1$ ( $f(r)=1+kr^{2}/L^{2}$) is the root of Eq. (%
\ref{Cubic}) for the vacuum spacetime ($m=0$ and $q=0$). The three solutions
of the cubic equation (\ref{Cubic}) are
\begin{align}
& \psi _{1}(r)=\frac{\tilde{\alpha}_{2}}{3\tilde{\alpha}_{3}}+\delta
+u\,\delta ^{-1},  \label{Sol1} \\
& \psi _{2}(r)=\frac{\tilde{\alpha}_{2}}{3\tilde{\alpha}_{3}}+-\frac{1}{2}%
(\delta +u\,\delta ^{-1})+i\frac{\sqrt{3}}{2}(\delta -u\,\delta ^{-1})
\label{Sol2} \\
& \psi _{3}(r)=\frac{\tilde{\alpha}_{2}}{3\tilde{\alpha}_{3}}-\frac{1}{2}%
(\delta +u\,\delta ^{-1})-i\frac{\sqrt{3}}{2}(\delta -u\,\delta ^{-1})
\label{Sol3}
\end{align}%
where
\begin{eqnarray*}
\delta &=&{(v+\sqrt{v^{2}-u^{3}})}^{1/3}, \\
u &=&\frac{\tilde{\alpha}_{2}^{2}-3\tilde{\alpha}_{3}}{9\tilde{\alpha}%
_{3}^{2}}, \\
v &=&\frac{1}{2}+\frac{2\tilde{\alpha}_{2}^{3}-9\tilde{\alpha}_{2}\tilde{%
\alpha}_{3}-27\tilde{\alpha}_{2}\tilde{\alpha}_{3}^{2}}{54\tilde{\alpha}%
_{3}^{3}}-\frac{1}{2\tilde{\alpha}_{3}}\left( \frac{mL^{2}}{r^{n}}-\frac{%
2L^{2}q^{2}}{(n-1)(n-2)r^{2(n-1)}}\right) .
\end{eqnarray*}%
All of the above three roots could be real in the appropriate range of $%
\tilde{\alpha}_{2}$ and $\tilde{\alpha}_{3}$. The second and third solutions
are real provided $u^{3}>v^{2}$. Here, we will consider only the first
solution $\psi _{1}(r)$ which is real provided $u^{3}<v^{2}$ at any $r$.

Now, we investigate the thermodynamics of the black hole solutions. The
temperature of the event horizon may be calculated through the use of
analytic continuation of the metric. One obtains
\begin{equation}
T=\frac{1}{4\pi \eta r_{+}}\left( [1-\tilde{\alpha}_{2}+\tilde{\alpha}_{3}]n%
\frac{r_{+}^{2}}{L^{2}}+k(n\mu -2\eta )-\frac{2q^{2}}{(n-1)r_{+}^{2(n-2)}}%
\right) ,  \label{Temp}
\end{equation}%
where%
\begin{equation}
\eta =1+2\tilde{\alpha}_{2}\frac{kL^{2}}{r_{+}^{2}}+3\tilde{\alpha}_{3}\frac{%
k^{2}L^{4}}{r_{+}^{4}}.  \label{eta}
\end{equation}%
The charge of the black holes per unit volume can be calculated by
integrating the flux of the electric field as
\begin{equation}
Q=\frac{1}{4\pi }\int^{\ast }{Fd\Omega }=\frac{1}{4\pi }q.
\end{equation}%
The electric potential $\Phi $, measured at infinity with respect to the
horizon is defined as
\begin{equation}
\Phi =A_{\mu }\chi ^{\mu }|_{r\rightarrow \infty }-A_{\mu }\chi ^{\mu
}|_{r=r_{+}}.  \label{electric potential}
\end{equation}%
where $\chi =\partial /\partial t$ is the null generator of the horizon. One
obtains
\begin{equation}
\Phi =\frac{q}{(n-2)r_{+}^{n-2}}.
\end{equation}

The finite total action in grand-canonical and canonical ensembles can be
found through the use of the counterterm method introduced in the last
section. It is a matter of straightforward calculations to show that the
total action is finite. Since we are interested in the stability of the
solutions in canonical ensemble, we calculate the finite Euclidian action
per unit volume $V_{k,n-1}$ in this ensemble. One obtains
\begin{eqnarray*}
\tilde{I} &=&\frac{\beta }{16\pi }\left\{\left( \frac{\xi }{(n-1)\eta }+%
\frac{1}{(n-2)}\right) \frac{2q^{2}}{r_{+}^{n-2}}+(n-1)k\mu r_{+}^{n-2}
\right. \\
&&\left.-(1-\tilde{\alpha}_{2}+\tilde{\alpha}_{3})\left[ n\left( \frac{\xi }{%
\eta }-1\right) +1\right] \frac{r_{+}^{n}}{L^{2}}-k(n\mu -2\eta )\frac{\xi }{%
\eta }r_{+}^{n-2}\right\}+I_{0}
\end{eqnarray*}%
where $\beta $ is the Euclidean time period (the inverse of temperature), $%
\eta $ is given in Eq. (\ref{eta}), $\xi $ is%
\begin{equation}
\xi =1+2\tilde{\alpha}_{2}\frac{(n-1)kL^{2}}{(n-3)r_{+}^{2}}+3\tilde{\alpha}%
_{3}\frac{(n-1)k^{2}L^{4}}{(n-5)r_{+}^{4}},  \label{xi}
\end{equation}%
and $I_{0}$ is
\begin{equation*}
I_{0}=\frac{\beta }{8\pi }\sum_{n/2=3,4,..}(-k)^{(n/2)}\frac{%
((n-1)!!)^{2}L^{n-2}}{n!}\left\{ 1+2\frac{(n-1)}{(n-3)}\tilde{\alpha}_{2}+3%
\frac{(n-1)}{(n-5)}\tilde{\alpha}_{3}\right\} .
\end{equation*}%
One should note that $I_{0}$ appears only in odd dimensions (even $n$).

For our static solution, there is a Killing vector field $\varsigma
=\partial /\partial t$ on the boundary, and therefore the quasilocal
conserved quantity associated with the stress tensors of Eq. (\ref{Stres})
is the mass of the black hole. Using the counterterm method introduced in
the last section, it is a matter of calculation to obtain the mass of black
hole as
\begin{eqnarray}
M &=&\frac{(n-1)m}{16\pi }+M_{0}  \notag \\
&=&\frac{(n-1)}{16\pi }\left\{ \left( 1-\tilde{\alpha}_{2}+\tilde{\alpha}%
_{3}+k\frac{L^{2}}{r_{+}^{2}}\mu \right) \frac{r_{+}^{n}}{L^{2}}+\frac{16\pi
^{2}Q^{2}}{(n-1)(n-2)r_{+}^{n-2}}\right\} +M_{0},  \label{Mass}
\end{eqnarray}%
where $M_{0}=I_{0}/\beta $ is the Casimir energy for the vacuum AdS metric
per volume $V_{k,n-1}$, which is nonzero only in odd dimensions. In order to
have positive energy $M-M_{0}$, one should restrict the range of Lovelock
coefficients in terms of $Q$, $L$ and $r_{+}$ as%
\begin{equation}
\tilde{\alpha}_{3}>-\left( 1+k^{3}\frac{l^{6}}{r_{+}^{6}}\right)
^{-1}\left\{ 1+k\frac{l^{2}}{r_{+}^{2}}+\tilde{\alpha}_{2}\left( 1-k^{2}%
\frac{l^{4}}{r_{+}^{4}}\right) -\frac{2q^{2}L^{2}}{(n-1)(n-2)r_{+}^{2(n-1)}}%
\right\} .  \label{alpha3}
\end{equation}%
That is, the third order Lovelock coefficient has a lower limit in terms of $%
\tilde{\alpha}_{2}$, $q$, $L$ and $r_{+}$.

Using the mass and action per unit volume $V_{k,n-1}$ calculated through the
use of counterterm method, one may calculate the entropy per unit volume
through the use of Gibbs-Duhem relation $S=\beta M-\tilde{I}$ as
\begin{equation}
S=\xi \frac{r_{+}^{n-1}}{4}.  \label{Ent}
\end{equation}%
It is worth to note that the entropy of the black hole solution per unit
volume $V_{k,n-1}$ calculated by the use of Gibbs-Duhem relation is
consistent with the calculation through the use of \ Wald formula \cite%
{Wald,Myers2}
\begin{equation}
S=\frac{1}{4}\int d^{n-1}x\sqrt{\tilde{g}}(1+kR+k^{2}\alpha _{2}\tilde{%
\mathcal{L}}_{2}),  \label{Wald}
\end{equation}%
where the integration is done on the $(n-1)$-dimensional spacelike
hypersurface of Killing horizon, $\tilde{g}_{\mu \nu }$ is the induced
metric on it, $\tilde{g}$ is the determinant of $\tilde{g}_{\mu \nu }$ and $%
\tilde{\mathcal{L}}_{2}$ is the $2$nd order Lovelock Lagrangian of $\tilde{g}%
_{\mu \nu }$. Also, one may note that the thermodynamic quantities
calculated in this section satisfy the first law of thermodynamics%
\begin{equation*}
dM=\frac{\left( \partial M/\partial r_{+}\right) _{Q}}{(\partial S/\partial
r_{+})}dS+\left( \frac{\partial M}{\partial Q}\right) _{S}dQ=TdS+\Phi dQ.
\end{equation*}

\section{Stability in the canonical ensemble\label{Stab}}

\begin{figure}[h]
{\includegraphics[width=.3\textwidth]{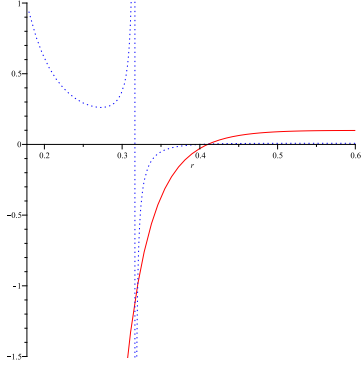}}
\caption{$10^{-2}T$ versus $r_{+}$ for $q=0.1$, $L=1$, $k=1$ and $n=6$ in
Einstein gravity (solid) and Lovelock gravity (dotted) with $\tilde{\protect%
\alpha}_{2}=0.1$ and $\tilde{\protect\alpha}_{3}=-0.01$. The vertical line
is $\protect\eta=0$ line.}
\label{Fig1}
\end{figure}

Now, we study the thermal stability of the black hole solutions of third
order Lovelock gravity in canonical ensemble. First, we investigate the
conditions of having black hole solution in Lovelock gravity. The solution
given by Eq. (\ref{Sol1}) presents a black hole solution provided $f(r)$ has
at least one real positive root. This occurs if $r_{+}\geq r_{ext}$, or $%
q\leq q_{ext}$ where $r_{ext}$ and $q_{ext}$ satisfy the following equation:
\begin{equation}
\lbrack 1-\tilde{\alpha}_{2}+\tilde{\alpha}_{3}]n\frac{r_{ext}^{2}}{L^{2}}%
+k[(n-2)+k(n-4)\frac{L^{2}}{r_{ext}^{2}}+k^{2}(n-6)\frac{L^{4}}{r_{ext}^{4}}-%
\frac{2q_{ext}^{2}}{(n-1)r_{ext}^{2(n-2)}}=0.  \label{ext}
\end{equation}
Also, the temperature of a physical black hole should be positive. One may
note that the temperature changes its sign at the root of $\eta=0$ ($%
r_{crit} $). The radius $r_{crit}$ depends on $L$ and Lovelock coefficients,
while $r_{ext}$ depends on $q$ too. In Fig. \ref{Fig1}, the vertical line is
$\eta=0 $ line. This figure shows that one may have only small ($\eta<0$)
and large black holes, and there is no medium black hole solution ($%
r_{crit}<r_{+}<r_{ext} $). This feature does not occur for black holes of
Einstein gravity or Lovelock gravity with positive Lovelock coefficients,
since $\eta \neq 0$. Thus, for the case of Lovelock gravity with negative
Lovelock coefficient(s) one may divide the black holes into two classes with
negative and positive $\eta$.

In the canonical ensemble $Q$ is fixed and therefore the black hole
solutions are stable provided $(\partial ^{2}M/\partial S^{2})_{Q}>0$ in the
range that $T$ is positive. Using the expressions for mass and entropy given
in Eq. (\ref{Mass}) and (\ref{Ent}), one may calculate $(\partial
^{2}M/\partial S^{2})_{Q}$:
\begin{eqnarray}
\left( \frac{\partial ^{2}M}{\partial S^{2}}\right) _{Q} &=&\left( \frac{%
\partial ^{2}M}{\partial r_{+}^{2}}\right) _{Q}\left( \frac{\partial S}{%
\partial r_{+}}\right) ^{-2}-\left( \frac{\partial M}{\partial r_{+}}\right)
_{Q}\left( \frac{\partial S}{\partial r_{+}}\right) ^{-3}\left( \frac{%
\partial ^{2}S}{\partial r_{+}^{2}}\right)  \notag \\
&=&\frac{(n-1)\sigma +2q^{2}\,\gamma r_{+}^{-2n+8}}{(n-1)^{2}r_{+}^{n+4}\eta
^{3}}.  \label{d2M}
\end{eqnarray}%
where
\begin{eqnarray*}
\gamma &=&(2n-3)+2k\tilde{\alpha}_{2}(2n-5)\frac{L^{2}}{r_{+}^{2}}+3k^{2}%
\tilde{\alpha}_{3}(2n-7)\frac{L^{4}}{r_{+}^{4}}, \\
\sigma&=&nr_{+}^{6}(1-\tilde{\alpha}_{2}+\tilde{\alpha}_{3})L^{-2}+kr_{+}^{4}%
\left[2+n(-1-6{\tilde{\alpha}_{2}}^2+ 6\tilde{\alpha}_{2}(1+\tilde{\alpha}%
_{3}))\right] \\
&&-k^{2}r_{+}^2\left[n(\tilde{\alpha}_{2}(15\tilde{\alpha}_{3}+1)-15\tilde{%
\alpha}_{3}(1+ \tilde{\alpha}_{3}))-8\tilde{\alpha}_{2}\right]L^2-2k\left[n({%
\tilde{\alpha}_{2}}^2-2\tilde{\alpha}_{3})- 4{\tilde{\alpha}_{2}}^2-6\tilde{%
\alpha}_{3}\right]L^4 \\
&&-3k^2\tilde{\alpha}_{3}\tilde{\alpha}_{2}r_{+}^{-2}(n-8)L^{6} -3k{\tilde{%
\alpha}_{3}}^2r_{+}^{-4}(n-6)L^8.
\end{eqnarray*}%
.
\begin{figure}[h]
{\includegraphics[width=.3\textwidth]{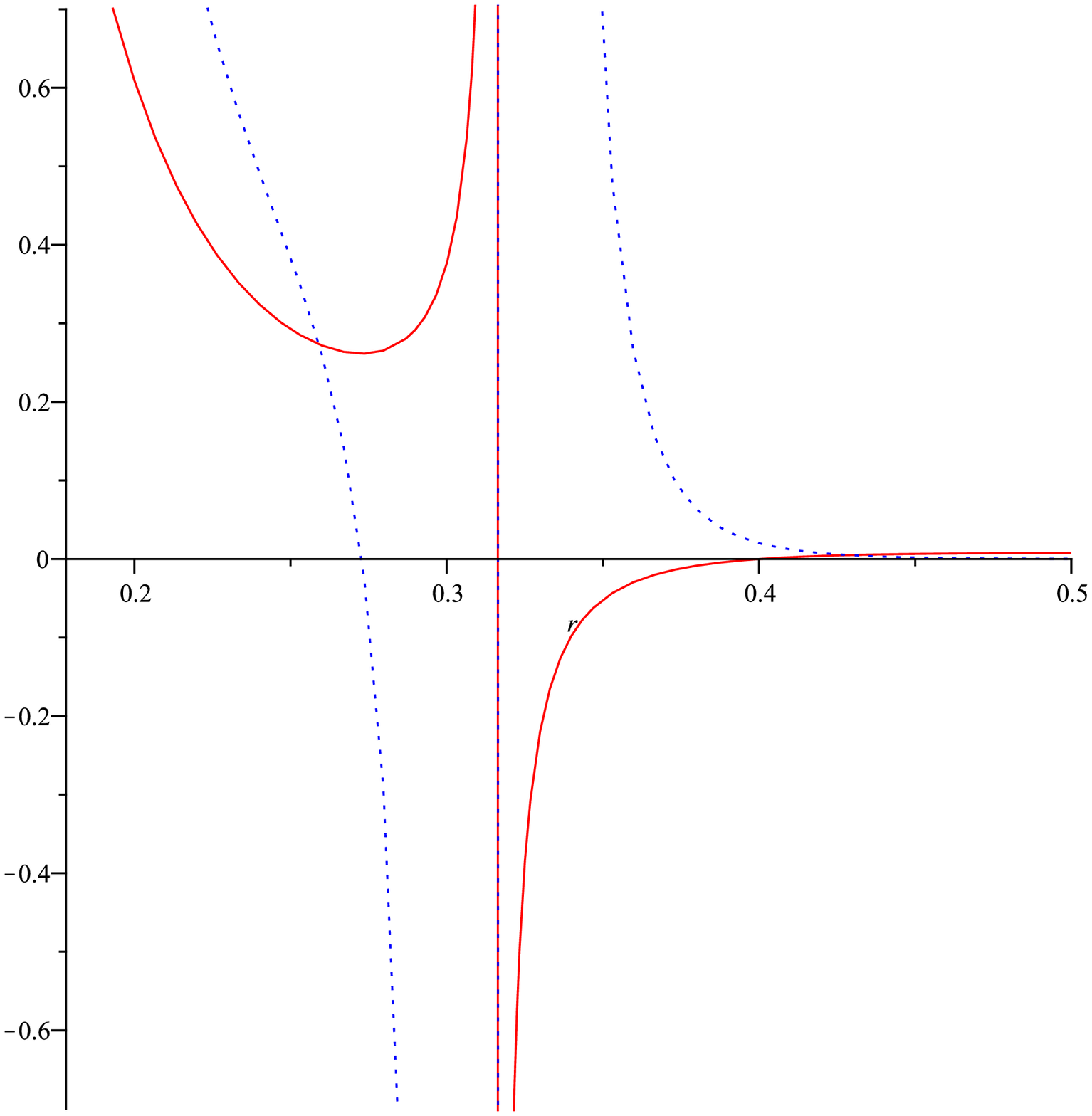}}
\caption{$10^{-3}T$ (dotted) and $10^{-5}(\partial ^{2}M/{\partial S^{2}})
_{Q}$ (solid) versus $r_{+}$ for $\tilde{\protect\alpha}_{2}=0.1,\tilde{%
\protect\alpha}_{3}=-0.01,q=0.1,L=1,k=1 $ and $n=6$. The vertical line is $%
\protect\eta=0$ line.}
\label{Fig2}
\end{figure}
\begin{figure}[h]
{\includegraphics[width=.3\textwidth]{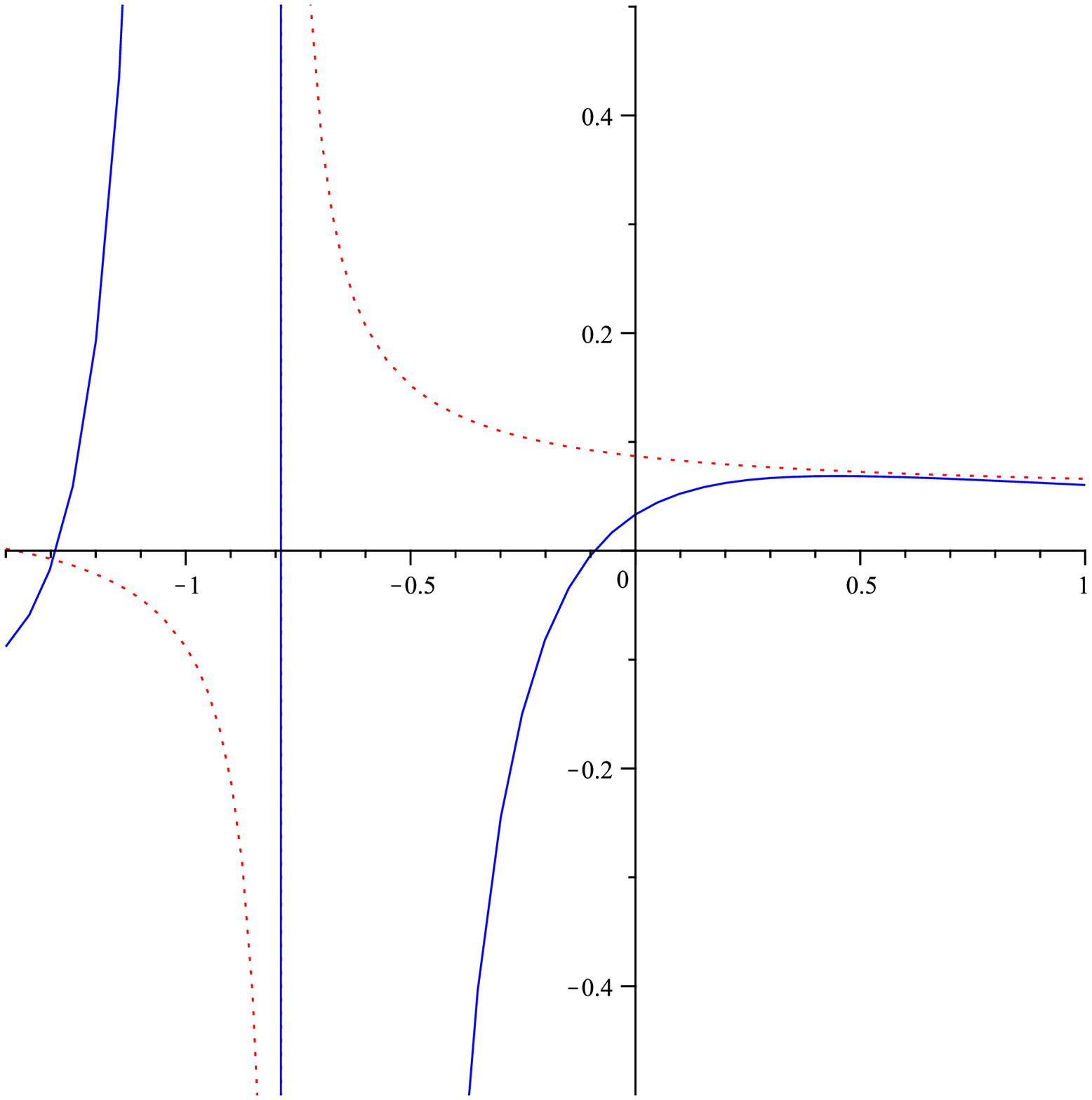}}
\caption{$10^{-2}T$ (dotted) and $10^{-1}\left(\partial ^{2}M/\partial
S^{2}\right) _{Q}$ (solid) versus $\tilde{\protect\alpha}_{3}$ for $\tilde{%
\protect\alpha}_{2}=0.1$, $q=0.5$, $L=1$, $k=1$, $n=6$ and $r_{+}=1.2$. The
vertical line is $\protect\eta=0$ line.}
\label{Fig3}
\end{figure}
\begin{figure}[h]
{\includegraphics[width=.3\textwidth]{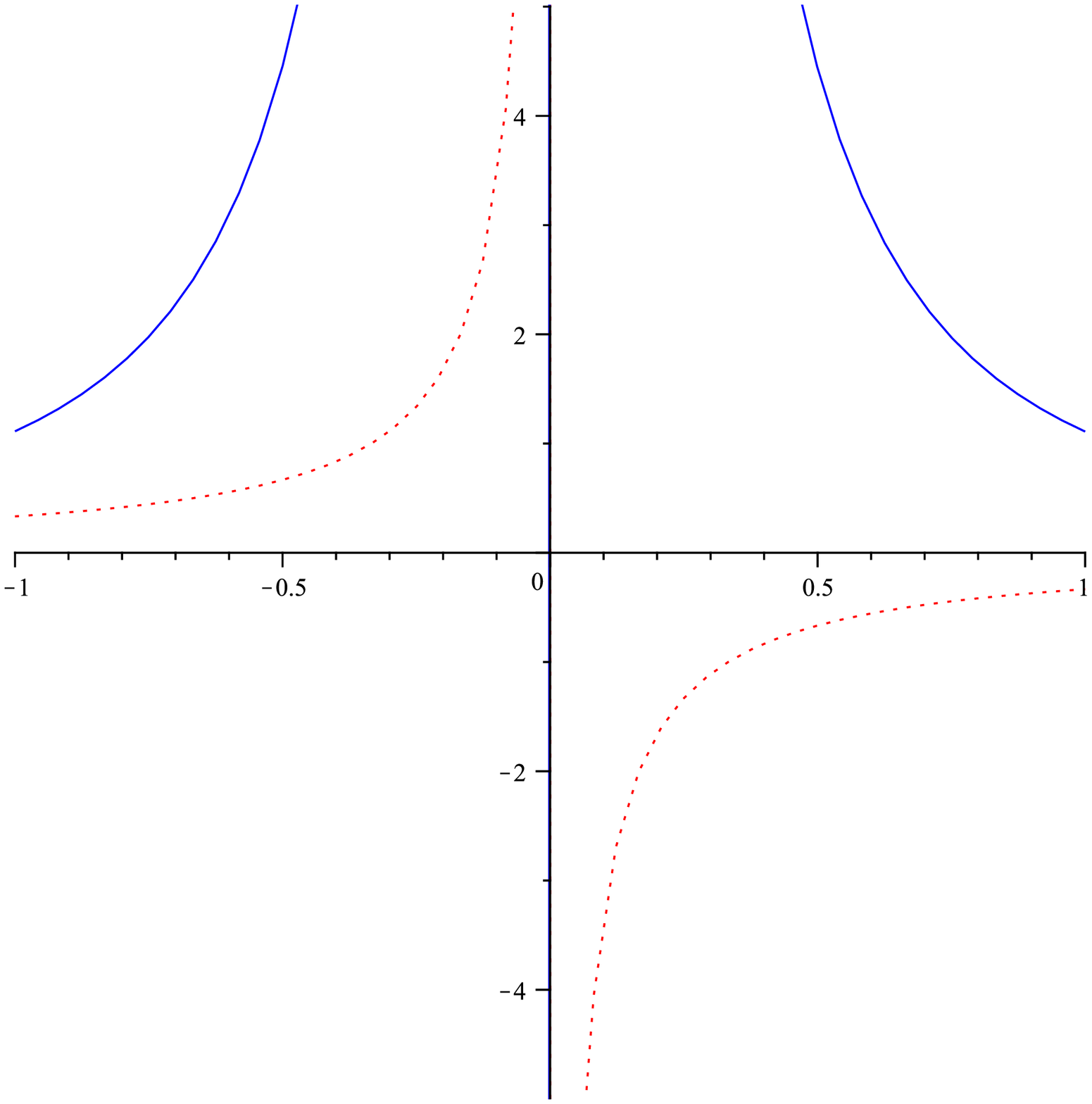}}
\caption{$10^{-4}T$ (dotted) and $10^{-4}\left(\partial ^{2}M/\partial
S^{2}\right) _{Q}$ (solid) versus $\tilde{\protect\alpha}_{3}$ for $\tilde{%
\protect\alpha}_{2}=0.1$, $q=0.5$, $L=1$, $k=1$, $n=6$ and $r_{+}=0.1$.}
\label{Fig4}
\end{figure}
\begin{figure}[h]
{\includegraphics[width=.3\textwidth]{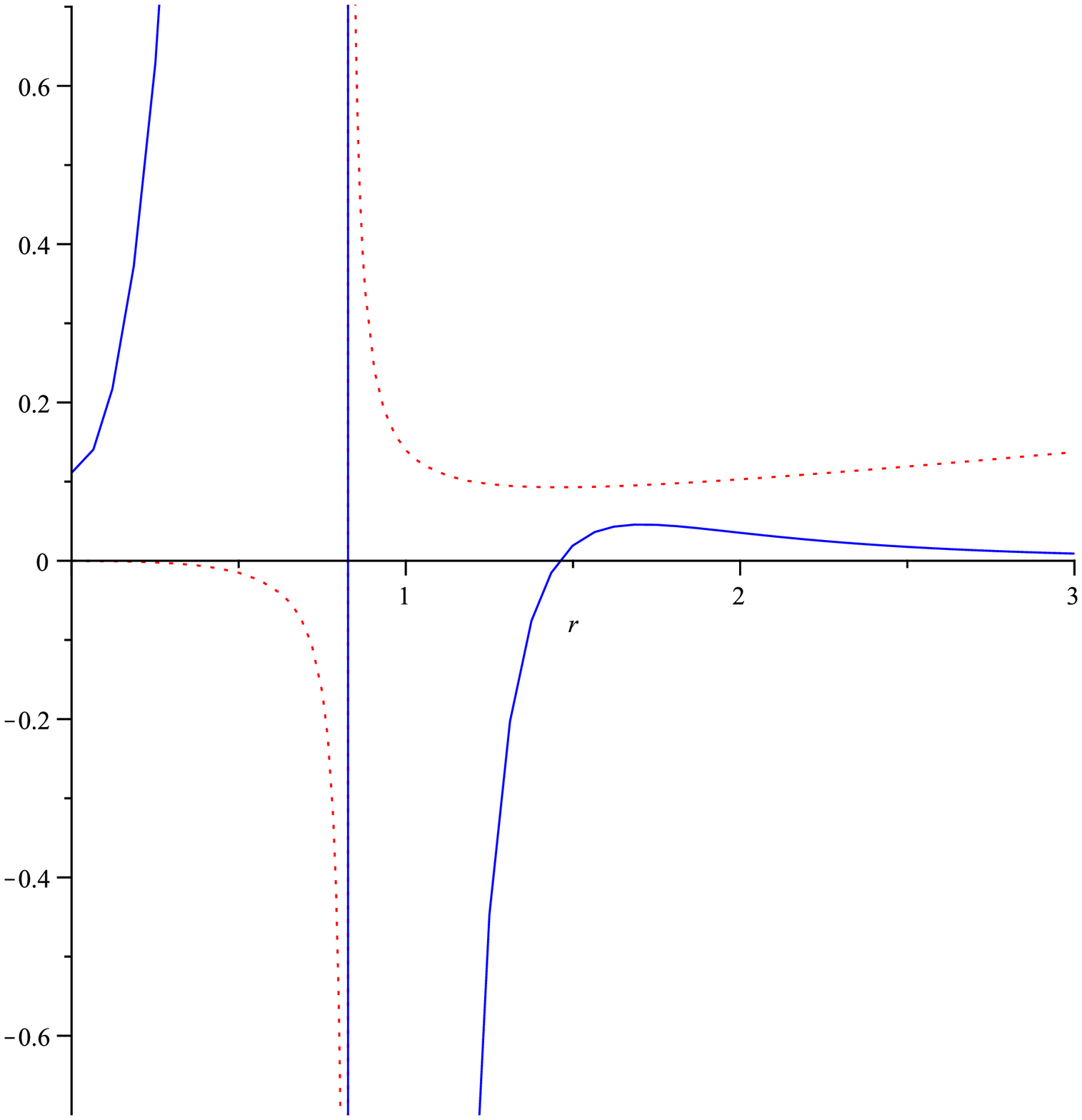}}
\caption{$10^{-2}T$ (dotted) and $\left(\partial ^{2}M/\partial S^{2}\right)
_{Q}$ (solid) versus $r_+$ for $\tilde{\protect\alpha}_{2}=0.1$, $\tilde{%
\protect\alpha}_{3}=-0.2$, $q=0$, $L=1$, $k=1$ and $n=6$. The vertical line
is $\protect\eta=0$ line.}
\label{Fig5}
\end{figure}
\begin{figure}[h]
{\includegraphics[width=.3\textwidth]{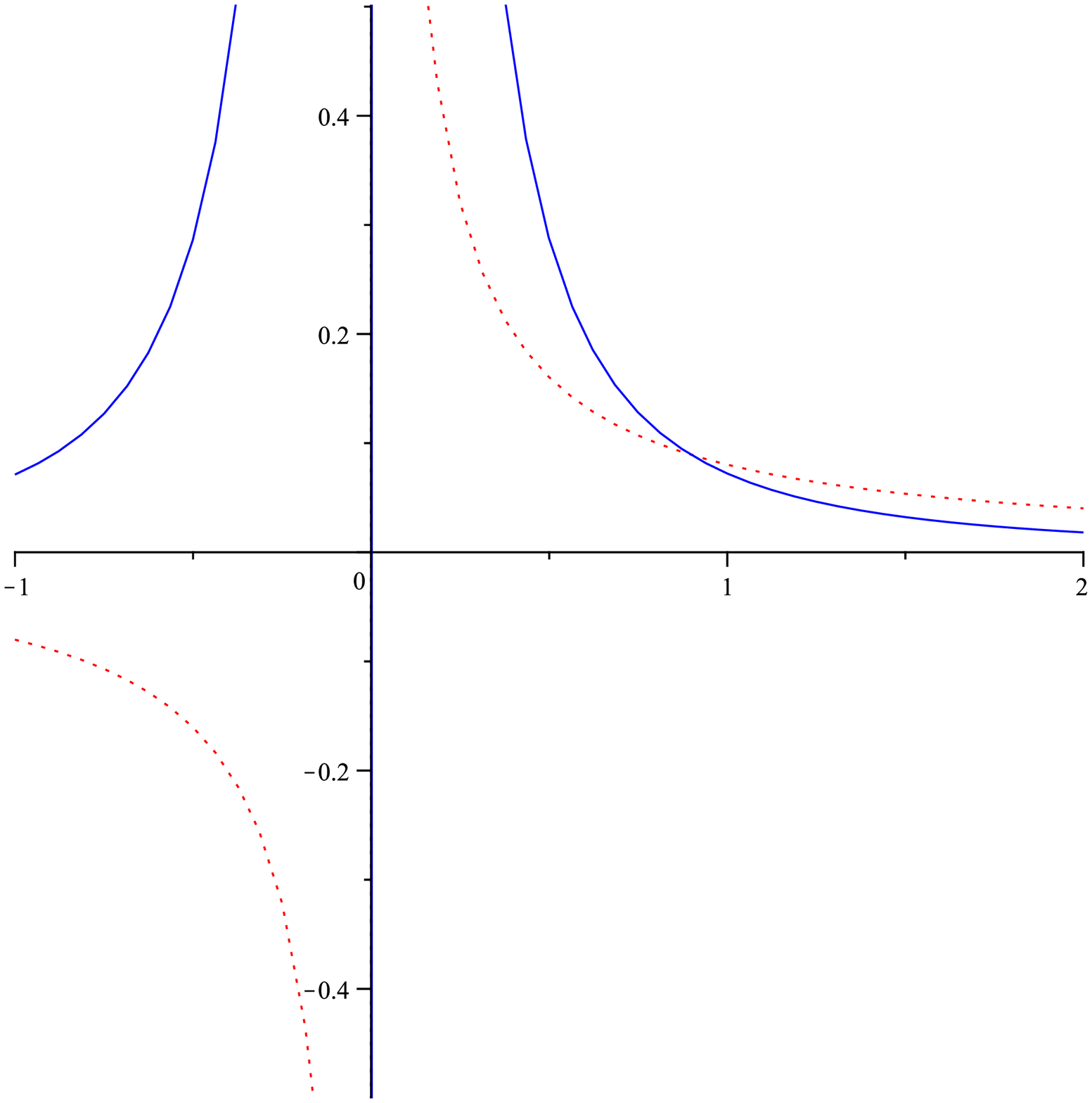}}
\caption{$10T$ (dotted) and $10\left(\partial ^{2}M/\partial S^{2}\right)
_{Q}$ (solid) versus $\tilde{\protect\alpha}_{3}$ for $\tilde{\protect\alpha}%
_{2}=0.1$, $q=0.5$, $L=1$, $k=1$, $n=6$ and $r_{+}=0.2$.}
\label{Fig6}
\end{figure}
\begin{figure}[h]
{\includegraphics[width=.3\textwidth]{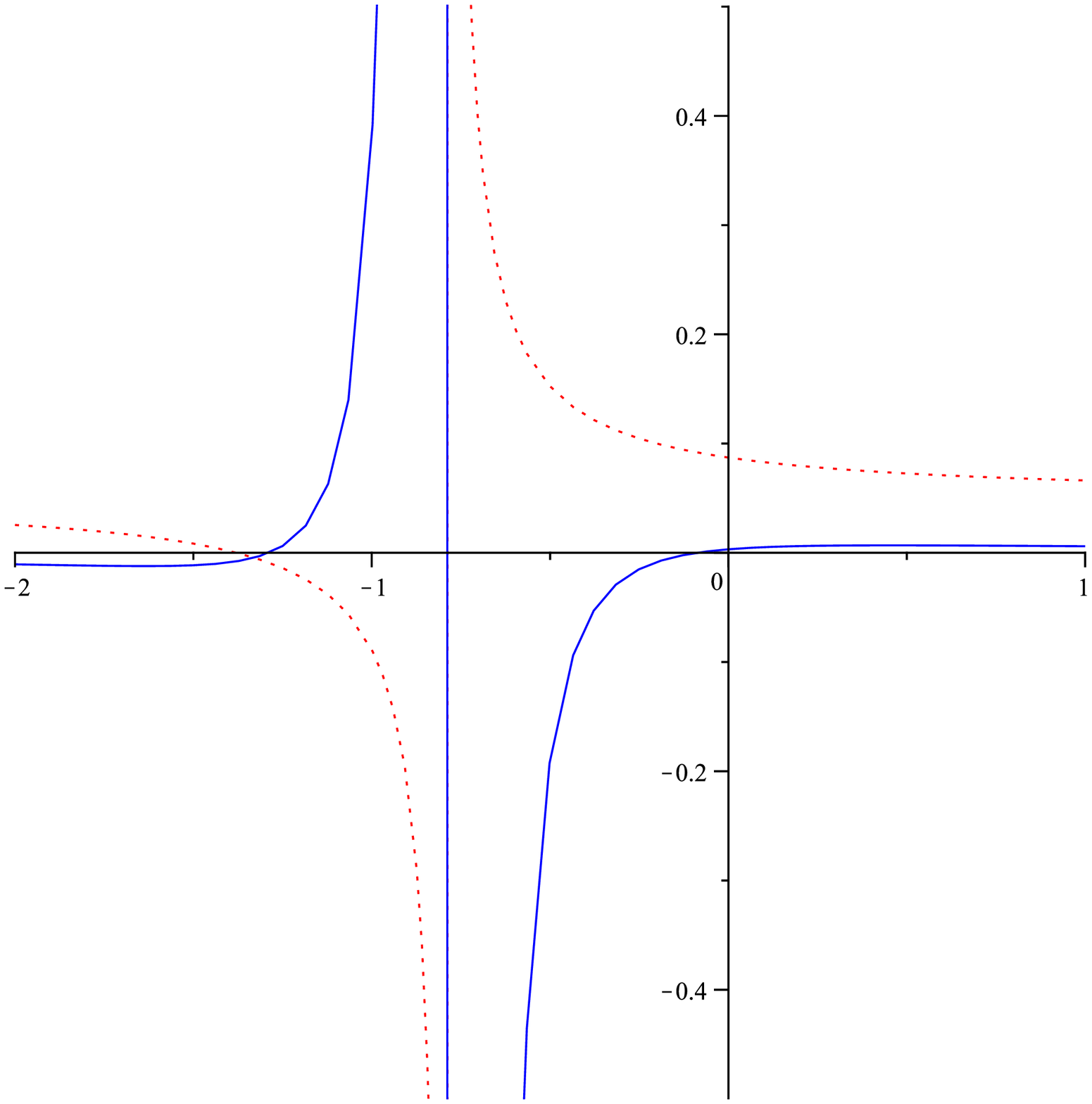}}
\caption{$10^{-2}T$ (dotted) and $10^{-2}\left(\partial ^{2}M/\partial
S^{2}\right) _{Q}$ (solid) versus $\tilde{\protect\alpha}_{3}$ for $\tilde{%
\protect\alpha}_{2}=0.1$, $q=0.5$, $L=1$, $k=1$, $n=6$ and $r_{+}=1.2$. The
vertical line is $\protect\eta=0$ line.}
\label{Fig7}
\end{figure}

To investigate the stability of black holes of third order Lovelock gravity
in canonical ensemble, we should find the sign of $\left( \partial
^{2}M/\partial S^{2}\right) _{Q}$, when $T$ is positive. In order to
investigate the stability, we plot $T$ and $\left(\partial ^{2}M/\partial
S^{2}\right) _{Q}$\ in one figure. The allowed region for investigation of $%
\left( \partial ^{2}M/\partial S^{2}\right) _{Q} $\ is when $T$ is positive.
In Fig. \ref{Fig2}, the vertical dotted-line is $\eta=0$ line. This figure
shows that the small black holes ($\eta<0$) divided into stable and unstable
black holes. That is a Hawking-Page phase transition exists for small black
holes between very small and small black holes. There is no phase transition
for large black holes ($\eta>0$) as one may see in Fig. \ref{Fig2}. For
investigating the effect of third order term of Lovelock gravity on the
stability of the black hole solutions, we plot $T$ and $\left( \partial
^{2}M/\partial S^{2}\right) _{Q}$ versus $\tilde{\alpha}_{3}$ for small and
large black holes. Figure \ref{Fig3} shows that there is no large black
holes for large negative $\tilde{\alpha}_{3}$, while the large black holes
for small negative $\tilde{\alpha}_{3}$ are not stable. As $\tilde{\alpha}%
_{3}$ becomes larger, large stable black holes may exist. Figure \ref{Fig4}
shows that there is no small black hole solution for positive $\tilde{\alpha}%
_{3}$, while the small black hole with negative $\tilde{\alpha}_{3}$ are
stable. Figure \ref{Fig5} shows that there is no negative-$\eta$ (small)
uncharged black hole with event horizon, while the large black holes ($%
\eta>0 $) show a phase transition between large and very large ones. One may
see in Fig. \ref{Fig5} that the large black holes with positive $\eta$ show
a phase transition, while these black holes with negative $\eta$ are not
stable. Finally, Figs. \ref{Fig6} and \ref{Fig7} show the temperature and $%
\left( \partial ^{2}M/\partial S^{2}\right) _{Q}$ versus $\tilde{\alpha}_{3}$
for small and large uncharged black holes. Figure \ref{Fig6} shows that
there is no uncharged black hole with negative $\eta$, while the black holes
with positive $\eta$ are stable. On the other hand, the large uncharged
black holes with positive $\eta$ show a phase transition as $\tilde{\alpha}%
_{3}$ becomes larger and these solutions are not stable for very large
negative $\tilde{\alpha}_{3}$.

\section{CLOSING REMARKS}

The concepts of action plays a central role in gravitation theories, but the
sum of the bulk and the surface terms diverges. In this paper, we introduced
the counterterms that remove the non-logarithmic divergences of
static solutions of third order Lovelock gravity. We did this by defining
the cosmological constant in such a way that the AdS metric (\ref{AdSmet})
is the vacuum solution of Lovelock gravity. Indeed, the cosmological
constant (\ref{Cos}) makes the asymptotic form of the solutions of Lovelock
gravity to be exactly the same as that of Einstein gravity. Thus, we
employed the counterterms introduced for Einstein gravity in \cite{Kraus}
and found that the power law divergences of static solutions in the
action of Lovelock gravity can be removed by suitable choice of
coefficients. We found that the counterterms are independent of Lovelock
coefficients and the dimensionally dependent of them is the same as those of
Einstein gravity. The main difference of our work with that of Ref. [16] is
that the counterterms are exactly the same as those of Einstein gravity and
do not depend on Lovelock coefficients. The only job which remains is that
one needs to calculate the coefficients $A_{p}$, $B_{p}$, $C_{p}$ and $D_{p}$%
. For example if one wants to remove the divergences of the action due to
the Gauss-Bonnet term, one should calculate $A_{2}$, $B_{2}$, $C_{2}$ and $%
D_{2}$ that remove the divergences of $\sqrt{\gamma }L_{2}$, without
regarding the other Lovelock terms. This enables one to generalize this
method to other higher curvature theories of gravity easily, including
fourth order Lovelock gravity or $f(R)$ gravity. We also introduced the
finite energy-momentum tensor in third order Lovelock gravity.

In addition, we employed these counterterms to calculate the finite action
and mass of the static black hole solutions of third order Lovelock
gravity. Calculating the temperature, the electric charge and electric
potential and using the calculated finite action and mass, we showed that
the entropy calculated through the use of Gibbs-Duhem relation is consistent
with the calculated entropy by Wald's formula. We, also, showed that the
conserved and thermodynamic quantities satisfy the first law of
thermodynamics. Finally, we investigated the stability of charged black
holes of Lovelock gravity in canonical ensemble. We found that the black
holes with respect to the sign of $\eta $ given in Eq. (\ref{eta}) may be
divided into two classes. The negative $\eta $ (small) black holes show a
phase transition between very small and small black holes, while the large
black holes are stable. There is no black hole solution with medium size. Of
course by small black holes we mean the size of them with respect to the
cosmological parameter $L$. We, also, investigated the effects of third
order Lovelock term on the stability of the solutions. We found that there
is no large black holes for large negative $\tilde{\alpha}_{3}$, while the
black holes for small negative $\tilde{\alpha}_{3}$ are not stable. We also
found that as $\tilde{\alpha}_{3}$ becomes larger, stable black holes may
exist. This shows that there is a phase transition as the third order term
of Lovelock gravity becomes larger. Finally, we found that there is no small
black hole solution for positive $\tilde{\alpha}_{3}$, while the black hole
with negative $\tilde{\alpha}_{3}$ are stable. Finally, we considered the
uncharged black holes of Lovelock gravity and found that negative $\eta $
solutions are not black holes with event horizon, while the positive $\eta $
black holes show a phase transition.

\acknowledgments{This work has been supported financially by Research
Institute for Astronomy \& Astrophysics of Maragha (RIAAM), Iran.}

\section{Appendix:}

The first term of Eq. (\ref{Stres}) which is the energy-momentum tensor of
the surface term is%
\begin{eqnarray*}
&&T_{ab}^{(sur)}=\frac{1}{8\pi }\{K_{ab}-K\gamma _{ab}+2\alpha
_{2}[3J_{ab}-J\gamma _{ab}-2\hat{G}_{(a}{}^{c}K_{b)c}+2\hat{R}_{ab}K-K_{ab}%
\hat{R}+2K_{cd}\hat{G}^{cd}\gamma _{ab}-2K^{cd}\hat{R}_{acbd}] \\
&&+3\alpha _{3}[5P_{ab}-P\gamma _{ab}+2K\hat{G}_{ab}^{(2)}+\mathcal{L}%
_{2}(K_{ab}+K\gamma _{ab})+4J\hat{R}_{ab}-24J_{(a}{}^{c}\hat{R}_{b)c}+8K^{cd}%
\hat{R}_{ac}\hat{R}_{bd}-8KK_{a}{}^{c}K_{b}{}^{d}\hat{R}_{cd} \\
&&+8KK_{ab}K^{cd}\hat{R}_{cd}-8K^{cd}\hat{R}_{ab}\hat{R}_{cd}+16K_{(a}{}^{c}%
\hat{R}_{b)}{}^{d}\hat{R}_{cd}+16K_{(a}{}^{c}K_{b)}{}^{d}K_{d}{}^{e}\hat{R}%
_{ce}-8K_{ab}K_{c}{}^{e}K^{cd}\hat{R}_{de}+6J_{ab}\hat{R} \\
&&-8K_{(a}{}^{c}\hat{R}_{b)c}\hat{R}-12J^{cd}\hat{R}_{acbd}-4K^{cd}\hat{R}%
\hat{R}_{acbd}-8K_{a}{}^{c}K_{bc}K^{de}\hat{R}%
_{de}+16K_{c}{}^{d}K^{ef}K_{(a}{}^{c}\hat{R}%
_{b)edf}+16K_{d}{}^{f}K^{de}K_{(a}{}^{c}\hat{R}_{b)ecf} \\
&&+16K_{(a}{}^{c}\hat{R}_{b)dce}\hat{R}^{de}-16KK^{de}K_{(a}{}^{c}\hat{R}%
_{b)dce}-16K^{cd}\hat{R}_{(a}{}^{e}\hat{R}_{b)cde}-4K^{cd}\hat{R}_{ac}{}^{ef}%
\hat{R}_{bdef}+8K^{cd}\hat{R}_{c}{}^{e}\hat{R}_{aebd} \\
&&+8K^{cd}\hat{R}_{c}{}^{e}\hat{R}_{adbe}-8K^{cd}\hat{R}%
_{a}{}^{e}{}_{c}{}^{f}\hat{R}_{bedf}-8K_{(a}{}^{c}\hat{R}_{b)}{}^{def}\hat{R}%
_{cdef}+8K_{a}{}^{c}K_{b}{}^{d}K^{ef}\hat{R}_{cedf}-4K_{ab}K^{cd}K^{ef}\hat{R%
}_{cedf} \\
&&+8K^{cd}\hat{R}_{a}{}^{e}{}_{b}{}^{f}\hat{R}_{cedf}+2\gamma _{ab}(\hat{G}%
_{cd}^{(2)}K^{cd}+6J^{cd}\hat{R}_{cd}-J\hat{R}+2KK^{cd}K^{ef}\hat{R}%
_{cedf}-4K_{c}{}^{e}K^{cd}K^{fh}\hat{R}_{dfeh})]\}
\end{eqnarray*}%
and the second term corresponding to the counterterm is
\begin{eqnarray*}
T_{ab}^{(ct)} &=&\frac{1}{8\pi }\sum_{p=1}^{3}\tilde{\alpha}_{p}\left\{
A_{p}\gamma _{ab}-2B_{p}\left( \hat{R}_{ab}-\frac{1}{2}\gamma \hat{R}\right)
\right.  \\
&&\left. -C_{p}\left[ -\,\gamma _{ab}\,\left( \hat{R}^{cd}\hat{R}_{cd}-\frac{%
n}{4(n-1)}\,\hat{R}^{2}\right) -\frac{n}{\,(n-1)}\,\hat{R}\hat{R}%
_{ab}\right. \right.  \\
&&\left. \left. -\frac{1}{(n-1)}\gamma _{ab}\,D^{2}\hat{R}+D^{2}\hat{R}_{ab}-%
\frac{(n-2)}{\,(n-1)}\,D_{a}\,D_{b}\hat{R}+\,4\hat{R}_{acbd}\hat{R}^{cd}%
\right] \right.  \\
&&\left. -D_{p}\left[ \frac{2(3n+2)}{4(n-1)}\left[ \left( \hat{G}_{ab}^{(1)}%
\hat{R}^{cd}\hat{R}_{cd}\right) -D_{a}D_{b}\left( \hat{R}^{ef}\hat{R}%
_{ef}\right) +\gamma _{ab}D^{2}\left( \hat{R}^{ef}\hat{R}_{ef}\right) +2\hat{%
R}\hat{R}_{a}^{\;c}\hat{R}_{bc}\right. \right. \right.  \\
&&\left. \left. +\gamma _{ab}D_{c}D_{d}\left( \hat{R}\hat{R}^{cd}\right)
+D^{2}\left( \hat{R}\hat{R}_{ab}\right) -D^{c}D_{b}\left( \hat{R}\hat{R}%
_{ac}\right) -D^{c}D_{a}\left( \hat{R}\hat{R}_{bc}\right) \right] -\frac{%
n(n+2)}{16(n-1)^{2}}\left[ -\gamma _{ab}\hat{R}^{3}\right. \right.  \\
&&\left. \left. +6\left( \hat{R}^{2}\hat{R}_{ab}-D_{a}D_{b}\hat{R}%
^{2}+\gamma _{ab}D^{2}\hat{R}^{2}\right) \right] -2\left[ -\gamma _{ab}\hat{R%
}^{ef}\hat{R}^{cd}\hat{R}_{ecfd}+3\left( \hat{R}_{a}^{\;e}\hat{R}^{cd}\hat{R}%
_{ecbd}+\hat{R}_{b}^{\;e}\hat{R}^{cd}\hat{R}_{ecad}\right) \right. \right.
\\
&&\left. -2D_{c}D_{d}\left( \hat{R}_{ab}\hat{R}^{cd}\right)
+2D_{c}D_{d}\left( \hat{R}_{a}^{\;c}\hat{R}_{b}^{\;d}\right) +2\gamma
_{ab}D_{e}D^{f}\left( \hat{R}^{cd}\hat{R}_{\;cfd}^{e}\right) +2D^{2}\left(
\hat{R}^{cd}\hat{R}_{acbd}\right) \right.  \\
&&\left. \left. -2D_{e}D_{a}\left( \hat{R}^{cd}\hat{R}_{\;cbd}^{e}\right)
-2D_{e}D_{b}\left( \hat{R}^{cd}\hat{R}_{\;cad}^{e}\right) \right] +\frac{n-2%
}{2(n-1)}\left[ -D_{a}\hat{R}D_{b}\hat{R}+\hat{R}_{b}{}^{c}D_{c}D_{a}\hat{R}%
\right. \right.  \\
&&\left. +\hat{R}_{a}{}^{c}D_{c}D_{b}(\hat{R})-2D_{c}D^{c}D_{b}D_{a}(\hat{R}%
)-\frac{7}{2}\hat{R}_{a}{}^{c}{}_{b}{}^{d}D_{c}D{d}(\hat{R})+2D_{a}(\hat{R}%
_{bc})D^{c}(\hat{R})+D_{b}(\hat{R}_{ac})D^{c}(\hat{R})\right.  \\
&&\left. +D_{b}\hat{R}_{ac}D^{c}\hat{R}-2D_{c}\hat{R}_{ab}D^{c}\hat{R}-\hat{R%
}_{b}{}^{c}{}_{c}{}^{d}D_{a}D_{d}\hat{R}-\hat{R}%
_{a}{}^{c}{}_{c}{}^{d}D_{d}D_{b}\hat{R}\right.  \\
&&\left. \left. -\hat{R}_{a}{}^{c}{}_{b}{}^{d}D_{d}D_{c}(\hat{R})+2\hat{R}%
_{ab}D_{e}D^{e}(\hat{R})+\frac{1}{2}\gamma _{ab}\left( D_{e}\hat{R}D^{e}\hat{%
R}+D^{4}\hat{R}\right) \right] \right.  \\
&&-\left. \left[ 2D_{a}D_{b}(D^{2}\hat{R})-2D^{4}\hat{R}_{ab}+2D_{p}\hat{R}%
_{(a}\hat{R}_{b)}^{p}-4\,\hat{R}^{pq}D_{a}D_{b}\hat{R}_{pq}+12\,\hat{R}%
^{pq}D_{(b}\hat{R}_{pa)q}-4(D^{2}\hat{R}^{pq})\hat{R}_{paqb}\right. \right.
\\
&&\left. +6D_{p}(\hat{R})D_{(b}\hat{R}_{a)}^{p}-2D_{a}(\hat{R}^{pq})D_{b}(%
\hat{R}_{pq})+4D_{(a}\,D_{q}\left( \hat{R}^{pq}\hat{R}_{b)p}\right)
+16D^{r}(\,\hat{R}^{pq})D_{(b}\hat{R}_{|rqp|a)}\right.  \\
&&\left. \left. -4\,\hat{R}^{pq}\hat{R}_{pa}\hat{R}_{qb}+8\,\hat{R}^{pr}\hat{%
R}_{r}^{q}\hat{R}_{paqb}+4\,\hat{R}^{pq}\hat{R}_{(a}^{r}\hat{R}_{|rqp|b)}%
\right] \right.  \\
&&+\left. \left. \gamma _{ab}\left[ -\,D^{4}\hat{R}-4\,D_{p}D_{q}(\hat{R})%
\hat{R}^{pq}+2\hat{R}_{pq}D^{2}(\hat{R}^{pq})+4\,D_{r}D_{s}(\hat{R}_{pq})%
\hat{R}^{prqs}-\,D_{e}(\hat{R})D^{e}(\hat{R})\right. \right. \right.  \\
&&\left. \left. +5\,D_{r}(\hat{R}_{pq})D^{r}(\hat{R}^{pq})-8\,D_{r}(\hat{R}%
_{pq})D^{q}(\hat{R}^{pr})-4\,\hat{R}_{pq}\hat{R}_{r}^{p}\hat{R}^{qr}+4\,\hat{%
R}_{pq}\hat{R}_{rs}\hat{R}^{prqs}\right] \right.  \\
&&\left. \left. +\frac{1}{n-1}\left[ 2\,D_{a}D_{b}(D^{2}\hat{R})-2\,(D^{2}%
\hat{R})\hat{R}_{ab}+D_{a}(\hat{R})D_{b}(\hat{R})+\gamma _{ab}\left(
-2\,D^{4}\hat{R}-\frac{1}{2}\,D_{c}\hat{R}D^{c}\hat{R}\right) \right] \right]
\right\} .
\end{eqnarray*}%
where $A_{p}$'s, $B_{p}$'s, $C_{p}$'s and $D_{p}$'s are given in Eqs. (\ref%
{A1}- \ref{A3}). The above counterterms remove the divergences of the
energy-momentum tensor for $n\leq 8$. As in the case of Einstein gravity,
one should add more counterterms for $n>8$.

\end{document}